# Title: Polarized x-rays from a magnetar

**Authors:** Roberto Taverna[1]*, Roberto Turolla[1,4], Fabio Muleri[2], Jeremy Heyl[3], Silvia Zane[4], Luca Baldini[5,6], Denis González-Caniulef[3], Matteo Bachetti[7], John Rankin[2], Ilaria Caiazzo[8], Niccolò Di Lalla[9], Victor Doroshenko[10], Manel Errando[11], Ephraim Gau[11], Demet Kırmızıbayrak[3], Henric Krawczynski[11], Michela Negro[12,13,14], Mason Ng[15], Nicola Omodei[9], Toru Tamagawa[16,17,18], Keisuke Uchiyama[17,18], Martin C. Weisskopf[19], Ivan Agudo[20], Lucio A. Antonelli[21,22], Wayne H. Baumgartner[19], Ronaldo Bellazzini[6], Stefano Bianchi[23], Stephen D. Bongiorno[19], Raffaella Bonino[24,25], Alessandro Brez[6], Niccolò Bucciantini[26,27,28], Fiamma Capitanio[2], Simone Castellano[6], Elisabetta Cavazzuti[29], Stefano Ciprini[30,22], Enrico Costa[2], Alessandra De Rosa[2], Ettore Del Monte[2], Laura Di Gesu[29], Alessandro Di Marco[2], Immacolata Donnarumma[29], Michal Dovčiak[31], Steven R. Ehlert[19], Teruaki Enoto[16], Yuri Evangelista[2], Sergio Fabiani[2], Riccardo Ferrazzoli[2], Javier A. Garcia[32], Shuichi Gunji[33], Kiyoshi Hayashida[34]†, Wataru Iwakiri[35], Svetlana G. Jorstad[36,37], Vladimir Karas[31], Takao Kitaguchi[16], Jeffery J. Kolodziejczak[19], Fabio La Monaca[2], Luca Latronico[24], Ioannis Liodakis[38], Simone Maldera[24], Alberto Manfreda[6], Frédéric Marin[39], Andrea Marinucci[29], Alan P. Marscher[36], Herman L. Marshall[15], Giorgio Matt[23], Ikuyuki Mitsuishi[40], Tsunefumi Mizuno[41], Stephen C.-Y. Ng[42], Stephen L. O'Dell[19], Chiara Oppedisano[24], Alessandro Papitto[21], George G. Pavlov[43], Abel L. Peirson[9], Matteo Perri[22,21], Melissa Pesce-Rollins[6], Maura Pilia[7], Andrea Possenti[7], Juri Poutanen[44,45], Simonetta Puccetti[22], Brian D. Ramsey[19], Ajay Ratheesh[2], Roger W. Romani[9], Carmelo Sgrò[6], Patrick Slane[46], Paolo Soffitta[2], Gloria Spandre[6], Fabrizio Tavecchio[47], Yuzuru Tawara[40], Allyn F. Tennant[19], Nicolas E. Thomas[19], Francesco Tombesi[48], Alessio Trois[7], Sergey Tsygankov[44,45], Jacco Vink[49], Kinwah Wu[4], Fei Xie[50]

**Affiliations:**

[1]Department of Physics and Astronomy, University of Padova; Via Marzolo 8, Padova, I-35131, Italy.

[2]INAF-IAPS, Roma; Via del Fosso del Cavaliere 100, I-00133 Roma, Italy.

[3]Department of Physics and Astronomy, University of British Columbia; Vancouver, BC V6T 1Z1, Canada.

[4]Mullard Space Science Laboratory, University College London; Holmbury St Mary, Dorking, Surrey RH5 6NT, UK.

[5]Università di Pisa, Dipartimento di Fisica Enrico Fermi; Largo B. Pontecorvo 3, I-56127 Pisa, Italy.

[6]Istituto Nazionale di Fisica Nucleare, Sezione di Pisa; Largo B. Pontecorvo 3, I-56127 Pisa, Italy.

[7]INAF-Osservatorio Astronomico di Cagliari; via della Scienza 5, I-09047 Selargius (CA), Italy.

[8]TAPIR, Walter Burke Institute for Theoretical Physics; Mail Code 350-17, Caltech, Pasadena, CA 91125, USA.

[9]W.W. Hansen Experimental Physics Laboratory, Kavli Institute for Particle Astrophysics and Cosmology, Department of Physics and SLAC National Accelerator Laboratory, Stanford University; Stanford, CA 94305, USA.

[10]Institut für Astronomie und Astrophysik, Universität Tübingen; Sand 1, 72076 Tübingen, Germany.

[11]Physics Department and McDonnell Center for the Space Sciences, Washington University in St. Louis; MO, 63130, USA.

[12]University of Maryland; Baltimore County, Baltimore, MD 21250, USA.

[13]NASA Goddard Space Flight Center; Greenbelt, MD 20771, USA.

[14]Center for Research and Exploration in Space Science and Technology, NASA/GSFC; Greenbelt, MD 20771, USA.

[15]MIT Kavli Institute for Astrophysics and Space Research, Massachusetts Institute of Technology; Cambridge, MA 02139, USA.

[16]RIKEN Cluster for Pioneering Research; 2-1 Hirosawa, Wako, Saitama 351-0198, Japan.

[17]RIKEN Nishina Center; 2-1 Hirosawa, Wako, Saitama 351-0198, Japan.

[18]Tokyo University of Science; 1-3 Kagurazaka, Shinjuku, Tokyo 162-8601, Japan.

[19]NASA/MSFC; Huntsville, AL, USA.

[20]Instituto de Astrofísica de Andalucía, IAA-CSIC; Glorieta de la Astronomía s/n, 18008 Granada, Spain.

[21]INAF Osservatorio Astronomico di Roma; Via Frascati 33, 00040 Monte Porzio Catone (RM), Italy.

[22]Space Science Data Center, Agenzia Spaziale Italiana; Via del Politecnico snc, 00133 Roma, Italy.

[23]Dipartimento di Matematica e Fisica, Università degli Studi Roma Tre; Via della Vasca Navale 84, 00146 Roma, Italy.

[24]Istituto Nazionale di Fisica Nucleare, Sezione di Torino; Via Pietro Giuria 1, 10125 Torino, Italy.

[25]Dipartimento di Fisica, Università degli Studi di Torino; Via Pietro Giuria 1, 10125 Torino, Italy.

[26]INAF Osservatorio Astrofisico di Arcetri; Largo Enrico Fermi 5, 50125 Firenze, Italy.

[27]Dipartimento di Fisica e Astronomia, Università degli Studi di Firenze; Via Sansone 1, 50019 Sesto Fiorentino (FI), Italy.

[28]Istituto Nazionale di Fisica Nucleare, Sezione di Firenze; Via Sansone 1, 50019 Sesto Fiorentino (FI), Italy.

[29]Agenzia Spaziale Italiana; Via del Politecnico snc, 00133 Roma, Italy.

[30]Istituto Nazionale di Fisica Nucleare, Sezione di Roma Tor Vergata; Via della Ricerca Scientifica 1, 00133 Roma, Italy.

[31]Astronomical Institute of the Czech Academy of Sciences; Boční II 1401/1, 14100 Praha 4, Czech Republic.

[32]California Institute of Technology; Pasadena, CA 91125, USA.

[33]Yamagata University;1-4-12 Kojirakawa-machi, Yamagata-shi 990-8560, Japan.

[34]Osaka University; 1-1 Yamadaoka, Suita, Osaka 565-0871, Japan.

[35]Department of Physics, Faculty of Science and Engineering, Chuo University; 1-13-27 Kasuga, Bunkyo-ku, Tokyo 112-8551, Japan.

[36]Institute for Astrophysical Research, Boston University; 725 Commonwealth Avenue, Boston, MA 02215, USA.

[37]Laboratory of Observational Astrophysics, St. Petersburg University; University Embankment 7/9, St. Petersburg 199034, Russia.

[38]Finnish Centre for Astronomy with ESO, University of Turku; Vesilinnantie 5, 20014 Turku, Finland.

[39]Université de Strasbourg, CNRS, Observatoire Astronomique de Strasbourg; UMR 7550, 67000 Strasbourg, France.

[40]Graduate School of Science, Division of Particle and Astrophysical Science, Nagoya University; Furo-cho, Chikusa-ku, Nagoya, Aichi 464-8602, Japan.

[41]Hiroshima Astrophysical Science Center, Hiroshima University; 1-3-1 Kagamiyama, Higashi-Hiroshima, Hiroshima 739-8526, Japan.

[42]Department of Physics, The University of Hong Kong; Pokfulam, Hong Kong.

[43]Department of Astronomy and Astrophysics, Pennsylvania State University; University Park, PA 16801, USA.

[44]Department of Physics and Astronomy, University of Turku; 20014 Turku, Finland.

[45]Space Research Institute of the Russian Academy of Sciences; Profsoyuznaya Str. 84/32, Moscow 117997, Russia.

[46]Center for Astrophysics, Harvard & Smithsonian; 60 Garden St, Cambridge, MA 02138, USA.

[47]INAF Osservatorio Astronomico di Brera; Via E. Bianchi 46, 23807 Merate (LC), Italy.

[48]Dipartimento di Fisica, Università degli Studi di Roma Tor Vergata; Via della Ricerca Scientifica, 00133 Roma, Italy.

[49]Anton Pannekoek Institute for Astronomy & GRAPPA, University of Amsterdam; Science Park 904, 1098 XH Amsterdam, The Netherlands.

[50]Guangxi Key Laboratory for Relativistic Astrophysics, School of Physical Science and Technology, Guangxi University; Nanning 530004, China.

∗Corresponding author. E-mail: taverna@pd.infn.it.

†Deceased

**Abstract:** We report on the first detection of linearly polarized x-ray emission from an ultra-magnetized neutron star with the Imaging X-ray Polarimetry Explorer (*IXPE*). The *IXPE* observations of the anomalous x-ray pulsar 4U 0142+61 reveal a linear polarization degree of $(12 \pm 1)\%$ throughout the *IXPE* 2–8 keV band. We detect a substantial variation of the polarization with energy: the degree is $(14 \pm 1)\%$ at 2–4 keV and $(41 \pm 7)\%$ at 5.5–8 keV, while it drops below the instrumental sensitivity around 4–5 keV, where the polarization angle swings by $\sim 90^\circ$. The *IXPE* observations give us completely new information about the properties of the neutron star surface and magnetosphere and lend further support to the presence of the quantum mechanical effect of vacuum birefringence.

**One-Sentence Summary:** The *IXPE* observation of 4U 0142+61 gives the first ever measurement of polarized emission from a magnetar in the x-rays.

**Main Text:** X-ray emission from magnetars is believed to be powered by the energy stored in their extremely strong magnetic fields (*1, 2*). There are currently close to 30 magnetar sources‡ (*3*), many of which are visible only during periods of enhanced activity (outbursts). A distinctive feature of magnetars is the emission of hard x-ray bursts, including the rare, hyperenergetic giant flares, spanning several orders of magnitude both in luminosity, $10^{38}$–$10^{47}$ erg s$^{-1}$, and duration, ≈ 0.1–100 s. Magnetars exhibit also pulsed emission at $L \approx 10^{33}$–$10^{35}$ erg s$^{-1}$ level, with spin periods $P \approx 2$–$12$ s and large spin-down rates, $\dot{P} \approx 10^{-14}$–$10^{-10}$ s s$^{-1}$; assuming a conventional spin-down model, this translates into magnetic fields up to $B \sim 10^{15}$ G (*4, 5*). The (soft) x-ray spectrum of magnetars is characterized by a blackbody (BB) component ($kT \sim 0.5$–1 keV) with a power-law (PL) tail extending to higher energies ($\approx 10$ keV). A second BB component in place of the PL tail provides a good fit for transient sources. Many objects are also detected in the hard x-rays (up to $\approx 200$ keV), where their spectrum can be approximated with a power law. Thermal emission is commonly attributed to radiation emitted by (different regions on) the cooling star surface, while the power law may originate from the up-scattering of thermal photons by charged particles flowing along the closed field lines of a "twisted" magnetosphere (*4, 5*).

Surface emission from highly magnetized neutron stars (NSs) is expected to be linearly polarized into two normal modes, the ordinary (O) and extraordinary (X) ones, with the polarization vector either parallel or perpendicular to the plane of the photon direction and the (local) magnetic field. The expected polarization degree strongly depends on the physical state of the outermost star layers. It is typically $\approx 10\%$ (or even less) if radiation comes from the bare, condensed surface while it can reach up to $\approx 80\%$ in the case of a magnetized atmosphere [see e.g. (*6–8*) and references therein]. As NSs are so small that they cannot be spatially resolved with current telescopes in any wavelength, their x-ray signal would exhibit a much lower linear polarization degree as the contributions from regions of the neutron star surface with different magnetic field orientations, and thus with different polarization orientations, would partially cancel each other (*9, 10*). On the other hand, for sufficiently strong magnetic fields the situation is drastically different, owing to a long predicted, but never measured as yet, effect of the quantum theory of electromagnetism (Quantum Electrodynamics, or QED). Strong magnetic fields, in fact, make the vacuum around the star birefringent, due to virtual electron-positron pairs of the quantum vacuum zipping in and out of existence (*11*). Vacuum birefringence forces the polarization vectors to follow the magnetic field direction and this results in an observed polarization degree much closer to that expected where radiation is emitted (*10, 12–14*). Characterizing the polarization properties of magnetar emission would therefore probe the physical conditions of the star surface and may even provide evidence for vacuum birefringence (*6–8*).

***IXPE* observations of 4U 0142+61:** The anomalous x-ray pulsar (AXP) 4U 0142+61 is among the brightest persistent magnetars with an (unabsorbed) flux of $\sim 7 \times 10^{-11}$ erg s$^{-1}$ cm$^{-2}$ in the 2–10 keV range, spin period $P = 8.69$ s and period derivative $\dot{P} = 2 \times 10^{-12}$ s s$^{-1}$. The spin-down magnetic field is $B \sim 1.3 \times 10^{14}$ G (*3, 15*).

Here we report on the polarimetric observation of 4U 0142+61 with the NASA/ASI mission Imaging X-ray Polarimetry Explorer [*IXPE*, (*16*)] carried out between 2022-01-31 and 2022-02-27 for a total of 840 ks. The *IXPE* Observatory provides imaging polarimetry over a nominal energy band of 2–8 keV. Data were extracted from Level 2 files and processed according to standard procedures [see the Supplementary Materials for further details].

Pulsations were clearly detected in the *IXPE* data with frequency $f = 0.115079332(8)$ Hz and frequency derivative $\dot{f} = -2.1(7) \times 10^{-14}$ Hz s$^{-1}$ (epoch MJD 59624.05054784674; errors are at 68.3% confidence level). These values are comparable with those reported in (*15*). We then carried out a spectral analysis of the source using XSPEC, version 12.12.1 (*17*). Single-component models provide a poor fit to the data, while a satisfactory agreement was found with several two-component models, either a blackbody plus power-law (BB+PL), a blackbody plus blackbody (BB+BB) or a blackbody plus a power-law truncated at low energies (below ~ 4 keV, BB+TPL); in all cases the value of the column density is largely unconstrained due to the lack of sensitivity below 2 keV. The parameters of the BB+PL fit and the (unabsorbed) flux ($\sim 7 \times 10^{-11}$ erg s$^{-1}$ cm$^{-2}$ in the 2–10 keV range) are in good agreement with those found by (*18*–*20*).

Calibrated event lists have been analyzed using the XPBIN tool inside IXPEOBSSIM, version 26.1.0 (*21*). This allowed us to extract the Stokes parameters $I$, $Q$ and $U$ of the events collected in each of the three detector units (DUs). After subtracting the background, the contributions of each DU were consistently added together, allowing for the fact that the three detectors are aligned at $120°$ with respect to each other and applying the proper calibration. Results for the phase-averaged normalized Stokes parameters $Q/I$ and $U/I$ in the 2–8 keV energy range are shown in Figure 1, together with those for the single DUs. The total (phase-averaged and energy-integrated) values are $Q/I = -0.013 \pm 0.008$ and $U/I = 0.120 \pm 0.008$, implying a polarization degree, PD = $\sqrt{Q^2 + U^2}/I$, of $(12 \pm 1)\%$ and a polarization angle, PA = $\arctan(U/Q)/2$, of $48° \pm 3°$, measured positive East of North; quoted errors are at $1\sigma$ level. The minimum detectable polarization (MDP) at the 99% confidence level for the 4U 0142+61 observation is $\sim 2\%$ over the 2–8 keV range and the significance of a non-zero polarization degree detection is at $\sim 15\sigma$ level. The same analysis was carried out independently using XSPEC obtaining consistent results (see Supplementary Materials).

In order to explore the behavior of PD and PA with energy, the data were grouped into 5 energy bins between 2 and 8 keV to ensure a sufficient number of counts in each bin. Results are shown in Figure 2 in the form of a polar plot where PD is the radial coordinate and PA the azimuth. The measured PD is $(14 \pm 1)\%$ at low energies (~ 2–4 keV), significantly above the MDP, which is ~ 4%. At 4–5 keV PD becomes consistent with zero, and then increases to $(41 \pm 7)\%$ in the last energy bin (5.5–8 keV), still above the MDP which is ~ 21% in this bin. The PA is roughly $50°$ at energies below 4 keV and swings by $90°$, to settle at $-40°$, above 5 keV.

We also performed a spectro-polarimetric fit by independently convolving the low- and the high-energy spectral components with a constant polarization model (POLCONST in XSPEC). The polarization angle displays a $90°$ swing for all the adopted spectral models. Moreover, for the BB+TPL model, the derived PD for the two components is in broad agreement with the measured one, with the BB less polarized than the high-energy PL (see Supplementary Materials).

A phase-dependent analysis was performed by sampling the flux with 100 phase bins and using the unbinned maximum likelihood technique outlined in (*22*) for PD and PA. The pulse profile (top panel of Figure 3) is double-peaked and resembles that reported in (*18*). Phase variations are clearly seen both in PD and in PA (middle and lower panels), with amplitudes $\sim 10\%$ and $\sim 30°$, respectively. The results are confirmed by a more detailed analysis with the IXPEOBSSIM and XSPEC software packages. Restricting the analysis to low energies (2–4 keV), we find that the main and secondary peak of the light curve exhibit a higher polarization fraction ($\sim$ 15%) compared to the valley ($\sim$ 9%). We also compared the phase dependent behavior of PA with the prediction of the rotating-vector model [RVM, (*23*)], again directly from the photon angle list and without binning in phase [see the Supplementary Materials for further details]. We found that a RVM with PA oscillating between 35° and 65° characterizes well the observed phase dependence of the polarization direction, although the geometric angles are poorly constrained.

We attempted a (preliminary) phase resolved spectral analysis of 4U 0142+61. We detected no statistically significant spectral changes with the rotational phase at this stage. In particular, the spectral parameters of the blackbody component are compatible with being constant in phase, in agreement with the analysis of (*24*) [see Figure S3].

**Discussion:** In this work we report on the first ever detection of x-ray polarization in a magnetar source, the AXP 4U 0142+61. The energy-resolved analysis revealed two distinct polarization patterns at low (2–4 keV) and high energies (5.5–8 keV). At $\sim$ 4–5 keV the polarization degree drops below the instrumental sensitivity (MDP $\sim$ 9%) and the polarization angle jumps from $48° \pm 2°$ at 2–4 keV to $-44° \pm 5°$ in the 5.5–8 keV energy band, providing robust evidence of a 90° swing. This indicates that photons are polarized in two different normal modes in the two energy bands.

The fact that the spectral shape of 4U 0142+61 does not significantly change with the rotational phase at low energies disfavors a scenario in which thermal radiation comes from a small region of the surface, e.g. a (point-like) polar cap. This is in apparent conflict with the finding that the polarization angle follows a rotating vector model. In fact, the RVM works only if the angle between a fixed direction and the projection of the local magnetic field in the plane of the sky is approximately the same for all the photons. This is not the case for emission from an extended region, since the direction of the (dipole) field changes substantially on the surface, unless one can find a way to lock the photon polarization vectors to the star magnetic field direction up to a large enough distance from the star, where the change in direction of the dipolar field is much smoother. This is precisely what vacuum birefringence is expected to do. QED effects, in fact, force the photon polarization vectors to align to the star magnetic field up to a distance, the polarization-limiting radius ($r_{\rm pl}$), which is typically $\sim$ 100 stellar radii for a magnetar (*10, 12–14*). As a result, on one hand the observed polarization degree turns out to be close to that at the emission. On the other hand, following the change in direction of the polarization vectors, the polarization angle changes continuously until photons arrive at $r_{\rm pl}$, where the polarization vectors freeze and the polarization angle does not change anymore [see the Supplementary Materials for more details]. This is in agreement with the observed phase-dependent behavior of PD and PA, with the former resembling the double-peaked profile of the flux, while the latter is reproduced by a RVM. For these reasons, we assume that vacuum birefringence does affect the polarization of the radiation in all the modelling discussed below, although the total polarization degree measured in 4U 0142+61 is not large enough to unambiguously prove that QED vacuum birefringence is indeed at work. In fact, for radiation coming from an extended surface region, detecting a phase- and energy-integrated polarization degree $\gtrsim$ 40% would have been required to firmly prove it [e.g. (*7*)].

The persistent emission from magnetars is often explained in terms of the Resonant Compton Scattering scenario [RCS; (*25*)], according to which thermal photons from the cooling star surface, which contribute to the soft x-ray emission, are up-scattered by charged particles flowing along the (closed) field lines of a twisted magnetosphere, giving rise to the high-energy power-law tail. Within the RCS model, photons in the high-energy tail of the x-ray spectrum should be polarized in the X-mode, with a polarization degree of $\sim 30\%$, quite independently on the polarization of primary radiation (*8*, *26–28*). The measured polarization degree in the 5.5–8 keV range, $\sim 35\%$, is suggestive that here X-mode photons dominate and, conversely, O-mode ones do so at low energies.

Although, at high energies, RCS can indeed produce a power-law tail polarized in the X-mode and with a PD close to the observed one, theoretical models presented so far for the surface emission predict, for the soft x-ray component, either a large polarization degree in the X-mode, $\gtrsim 50\%$ in the case of a gaseous atmosphere heated from below (*29–32*), or a small polarization degree in the O-mode, $\lesssim 10\%$ for a condensed surface (*29*, *33–35*) [see also (*6–8*)]. These models assume that emission comes from the entire surface of the neutron star; however, according to the results of our phase-resolved spectral analysis, we have evidence that radiation should come from a rather extended region of the star surface. Recent 3D simulations of magneto-thermal evolution indicate that a hotter belt may appear on the star surface, close to the magnetic equator (*36*). We find that radiation from an iron condensed surface emitted from an equatorial belt results in a predominance of O-mode photons at low energies (2–4 keV) with PD $\sim 10$–$15\%$. Reprocessing by RCS then produces an excess of X-mode photons at higher energies (5.5–8 keV) with PD $\sim 35$–$40\%$. The stars in Figure 2 show the results for such an emission model, assuming a magnetic field strength $\sim 10^{14}$ G, as derived for 4U 0142+61 (*18*), and the fixed-ion approximation for the magnetized iron condensate (*34*,*35*). The broad quantitative agreement of the modeled and observed polarization fraction and direction supports this scenario.

A further possible mechanism to produce an O-mode dominated signal at low energies is through emission by a gaseous layer with an “inverted temperature” structure. In fact, the reason why NS atmospheric cooling models predict a signal which is X-mode dominated is that, typically, the temperature of the atmosphere decreases with decreasing optical depth. X-mode photons, which have a smaller cross-section, decouple in a deeper, hotter layer and dominate the flux. In this respect, one may speculate that, if for some reason the temperature gradient is inverted (decreasing inward) near the photospheric region, then O-mode photons, which originate from shallower layers of the atmosphere, would decouple in a hotter region and dominate the outgoing flux. Such an “inverted” temperature profile can be produced, e.g., if there is a significant downward flow of energy, as in the presence of an external particle bombardment. Simulations of neutron star atmospheres with particle bombardment show indeed a shallow inverted temperature gradient (*37*). O-mode emission at the surface in the context of a condensed Fe surface or an atmosphere with inverted temperature gradient (as an externally illuminated one), followed by RCS magnetospheric reprocessing can therefore provide a consistent (although not unique) explanation for the observations.

In a completely alternative scenario, one could interpret the low-energy emission as polarized in the X-mode and, consequently, the one above the observed mode-shifting at 4–5 keV in the O-mode. Low-energy, X-mode dominated emission with a low polarization degree ($\sim 15\%$) may still originate from an extended region of a Fe condensed surface, for example by looking at the emitting equatorial belt previously discussed from above the magnetic equator or replacing the equatorial belt with a polar cap and observing it from above the pole. Radiation from a thin atmosphere/corona in the presence of unsaturated thermal Comptonization (*8*) may also produce the same polarization pattern. In this picture, however, it is more difficult to explain how O-mode photons can dominate the emission in the 5–8 keV band. In principle, this could be realized invoking saturated Comptonization in a thin atmosphere/corona (*8*) or through emission from a pair plasma (*38*), but all these models predict a PD much higher than the observed one. Emission from a small region of the surface covered by an externally illuminated gaseous layer (as discussed before) and hot enough to dominate the high-energy band will also result in substantial polarization in the O-mode. For this alternative picture, however, no detailed models are presently available to provide a quantitative expectation of the observed spectral and polarization properties.

Knowledge of the mode (either O or X) in which the observed x-ray photons are predominantly polarized allows us to estimate the orientation of the projection of the star spin axis in the plane of the sky. It can be shown, in fact, that the phase-averaged PA is $0^\circ$ ($90^\circ$) for radiation mostly polarized in the O-mode (X-mode) when the reference direction in the plane of the sky is along the spin axis projection [see e.g. (*6*, *10*)]. Assuming that radiation in the 2–4 keV range is polarized in the O-mode with PA $\sim 50^\circ$ implies that the projection of the spin axis is at an angle $\sim 50^\circ$ East of North. Conversely, if low-energy photons are polarized in the X-mode the spin axis projection lies at $\sim 40^\circ$ West of North. In the latter case the spin projection would be broadly consistent with the direction of the magnetar proper motion, $60^\circ \pm 12^\circ$ West of North (*39*) [see Figure 2], while in the former the two would be almost orthogonal. The spin-proper motion alignment/disalignment in pulsars is a long-standing puzzle. Observational evidences argue in favor of the alignment for Crab and Vela, as well as for a number of other pulsars [at least with a not too large kinematic age, see (*40*) and references therein]. On the other hand, as discussed in (*41*), perpendicular kicks may result from the evolution of evanescent binary systems. In this respect, future x-ray polarization measurements, carried out systematically in several (radio-silent) neutron stars, might help to better understand this issue, in this way shedding light on the processes which occur during the acceleration phase mechanism in the proto-NS formation stage and on the subsequent interaction of the star kick with the Galactic gravitational potential.

The *IXPE* discovery of two competing polarization modes in 4U 0142+61 shows that x-ray polarimetry opens up our models of the elusive processes of the polarized emission from neutron star surfaces and atmospheres, mode propagation and mode switching in extreme magnetic fields, resonant cyclotron scattering and QED vacuum birefringence to detailed observational tests. The scenario in which radiation is emitted from an extended region of the stellar condensed surface within the RCS model provided a plausible explanation of the *IXPE* polarization measurement. Spectro-polarimetric observations of this and similar sources (such as the AXP 1RXS J170849.0-400910 that *IXPE* will observe later this year) will likely allow us to test and refine the geometry of the emission region(s) and the models themselves, and to arrive at a firmer identification of the modes dominating the emission in different energy bands.

## References and Notes

‡ See the on-line catalog at http://www.physics.mcgill.ca/ pulsar/magnetar/main.html.

**Acknowledgments:**

**Funding:**

**Author contributions:**

**Competing interests:** Authors declare that they have no competing interests.

**Data and materials availability:** Data of the 4U 0142+61 observation are available in the HEASARC IXPE Data Archive (https://heasarc.gsfc.nasa.gov/docs/ixpe/archive/).

**Supplementary Materials**

Materials and Methods

Figs. S1 to S6

Tables S1 to S3

References (*43–65*)

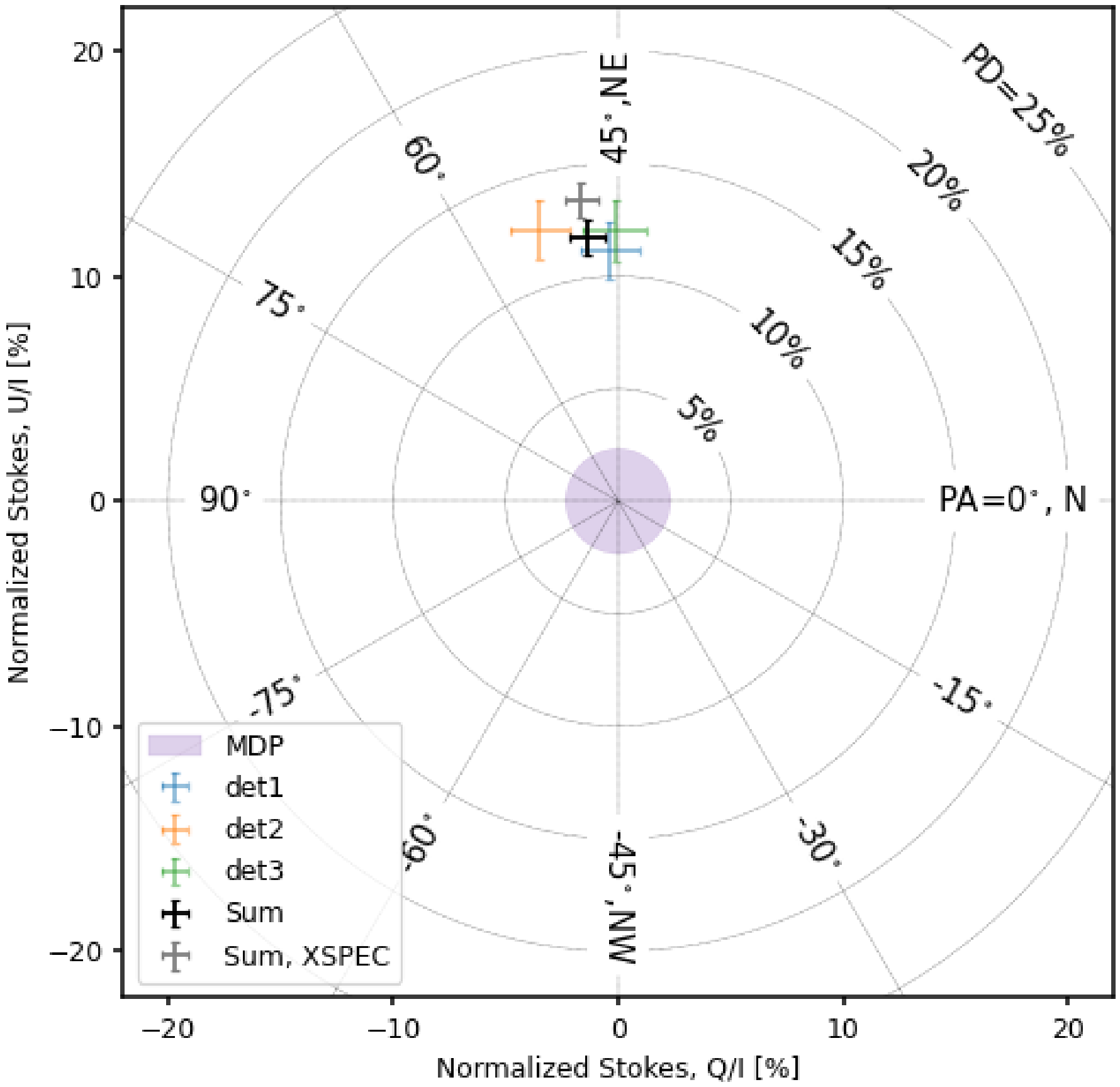

**Fig. 1. The normalized Stokes parameters $Q/I$ and $U/I$ measured by the three *IXPE* detector units (colored points with error bars), together with the total values obtained with the model-independent approach described in (*42*) [black] and with XSPEC [gray, see the SM for details].** The adopted energy interval is 2–8 keV. Circles centered at the origin give the contours of constant PD, whereas radial lines those of constant PA. The shaded area shows the MDP at the 99% confidence level for the three summed measurements. Data were background-subtracted and processed with the IXPEOBSSIM suite.

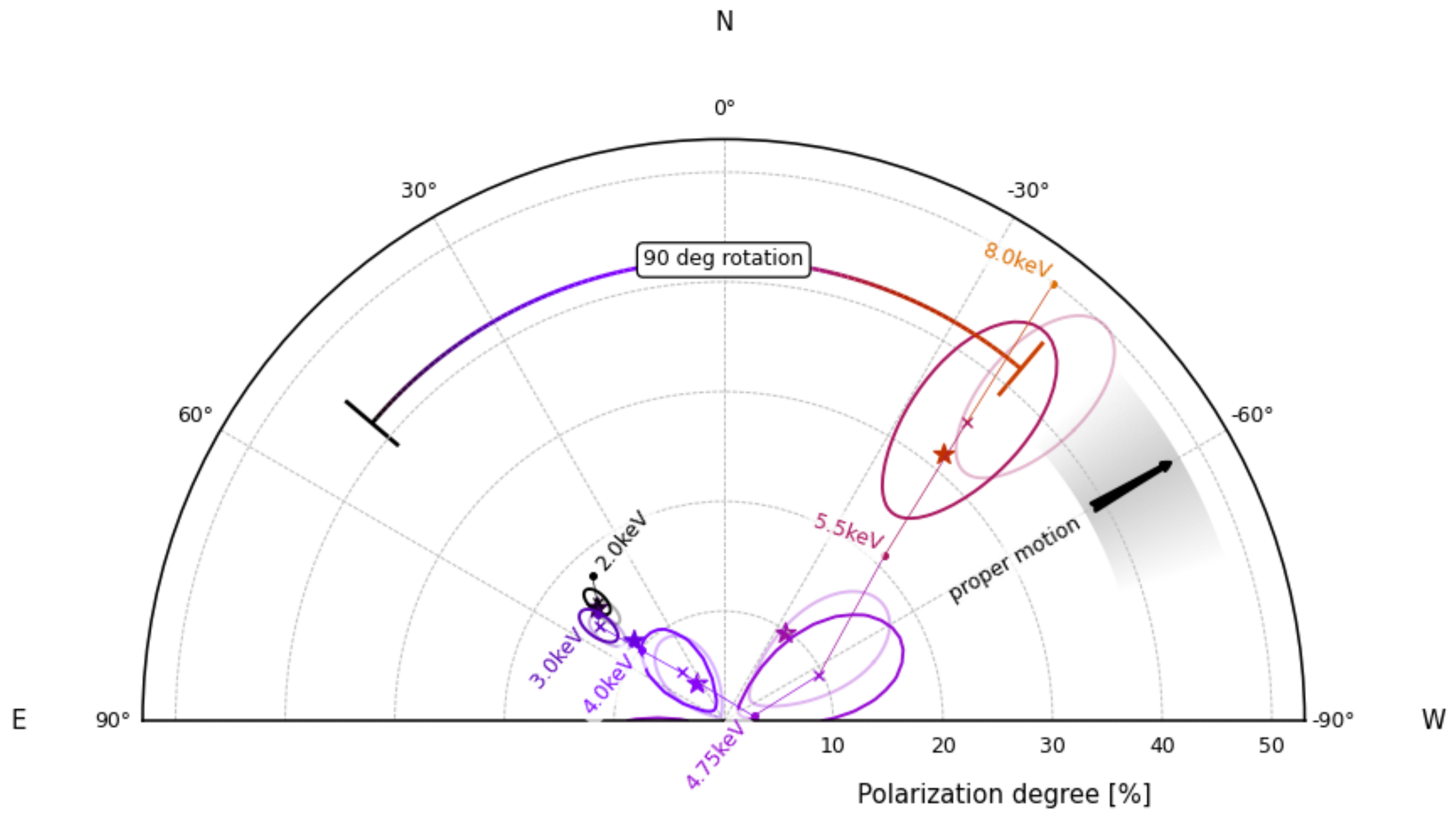

**Fig. 2. Polar plot showing the measured PD (radius) and PA (azimuth) at different energies in the 2–8 keV band.** Contours enclose the 68.3% confidence level regions obtained with the model-independent approach described in (*42*) [thin lines] and with XSPEC [thick lines, see the SM for details]. The arrow and the shaded area indicate the proper motion direction of the source and the associated uncertainty (*39*). The results of a condensed-surface RCS model for PD and PA in the same energy bins are shown by stars (see Discussion).

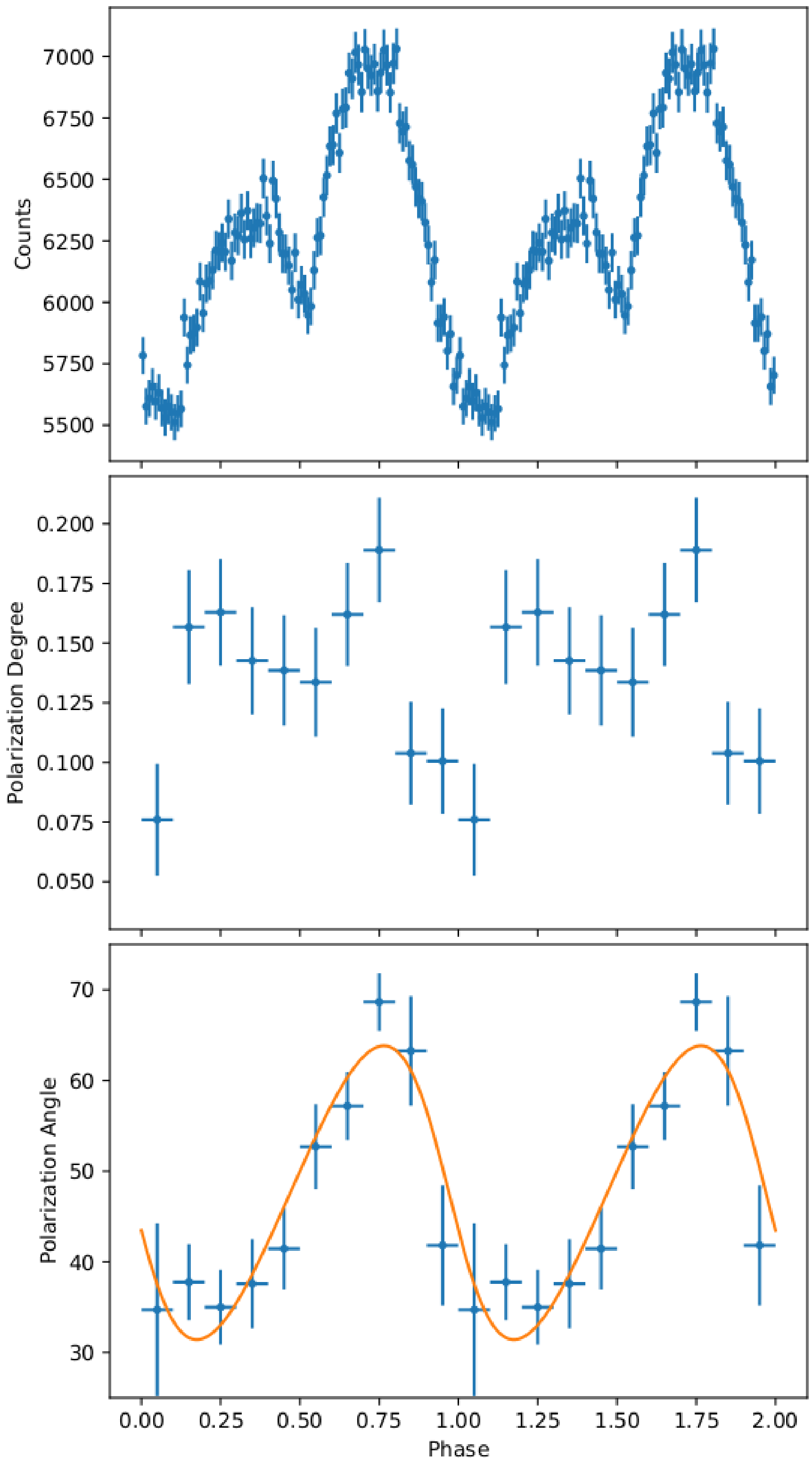

**Fig 3. The energy-integrated (2–8 keV range) counts (upper), PD (center) and PA (lower).** The uncertainties in the middle and lower panels correspond to $\Delta \log L = 1$ contours of the unbinned likelihood. The orange curve in the lowermost panel shows the best-fitting rotating vector model for the polarization angle.

# Supplementary Materials for

Polarized x-rays from a magnetar

Roberto Taverna[1]*, Roberto Turolla[1,4], Fabio Muleri[2], Jeremy Heyl[3], Silvia Zane[4], Luca Baldini[5,6], Denis González-Caniulef[3], Matteo Bachetti[7], John Rankin[2], Ilaria Caiazzo[8], Niccolò Di Lalla[9], Victor Doroshenko[10], Manel Errando[11], Ephraim Gau[11], Demet Kırmızıbayrak[3], Henric Krawczynski[11], Michela Negro[12,13,14], Mason Ng[15], Nicola Omodei[9], Toru Tamagawa[16,17,18], Keisuke Uchiyama[17,18], Martin C. Weisskopf[19], Ivan Agudo[20], Lucio A. Antonelli[21,22], Wayne H. Baumgartner[19], Ronaldo Bellazzini[6], Stefano Bianchi[23], Stephen D. Bongiorno[19], Raffaella Bonino[24,25], Alessandro Brez[6], Niccolò Bucciantini[26,27,28], Fiamma Capitanio[2], Simone Castellano[6], Elisabetta Cavazzuti[29], Stefano Ciprini[30,22], Enrico Costa[2], Alessandra De Rosa[2], Ettore Del Monte[2], Laura Di Gesu[29], Alessandro Di Marco[2], Immacolata Donnarumma[29], Michal Dovčiak[31], Steven R. Ehlert[19], Teruaki Enoto[16], Yuri Evangelista[2], Sergio Fabiani[2], Riccardo Ferrazzoli[2], Javier A. Garcia[32], Shuichi Gunji[33], Kiyoshi Hayashida[34]†, Wataru Iwakiri[35], Svetlana G. Jorstad[36,37], Vladimir Karas[31], Takao Kitaguchi[16], Jeffery J. Kolodziejczak[19], Fabio La Monaca[2], Luca Latronico[24], Ioannis Liodakis[38], Simone Maldera[24], Alberto Manfreda[6], Frédéric Marin[39], Andrea Marinucci[29], Alan P. Marscher[36], Herman L. Marshall[15], Giorgio Matt[23], Ikuyuki Mitsuishi[40], Tsunefumi Mizuno[41], Stephen C.-Y. Ng[42], Stephen L. O'Dell[19], Chiara Oppedisano[24], Alessandro Papitto[21], George G. Pavlov[43], Abel L. Peirson[9], Matteo Perri[22,21], Melissa Pesce-Rollins[6], Maura Pilia[7], Andrea Possenti[7], Juri Poutanen[44,45], Simonetta Puccetti[22], Brian D. Ramsey[19], Ajay Ratheesh[2], Roger W. Romani[9], Carmelo Sgrò[6], Patrick Slane[46], Paolo Soffitta[2], Gloria Spandre[6], Fabrizio Tavecchio[47], Yuzuru Tawara[40], Allyn F. Tennant[19], Nicolas E. Thomas[19], Francesco Tombesi[48], Alessio Trois[7], Sergey Tsygankov[44,45], Jacco Vink[49], Kinwah Wu[4], Fei Xie[50]

Correspondence to: taverna@pd.infn.it

**This PDF file includes:**

Materials and Methods
Figs. S1 to S6
Tables S1 to S3

## Materials and Methods

Observations and data analysis

The *IXPE* observatory includes three identical x-ray telescopes, each comprising an x-ray mirror assembly and a polarization-sensitive pixelated detector (*43*, *44*), for measuring the energy, arrival direction, arrival time and linear polarization of the detected x-ray signal. The data can be analyzed with either the standard softwares or, especially for polarimetry, which is a distinctive feature of *IXPE*, with the custom software developed by the *IXPE* collaboration collected in a suite named IXPEOBSSIM (*21*). To validate the results, we performed the polarization analysis independently with both *IXPE* custom tools and XSPEC (*17*), using the appropriate instrument response functions provided in the public *IXPE* calDB, as detailed in the following. We anticipate that the results were compatible within statistical uncertainties.

Data analysis started from Level 2 files available in the *IXPE* archive at HEASARC§. We joined the two 4U 0142+61 observations available at the time of writing: the first started on 2022-01-31 at 07:37:07 UTC and ended on 2022-02-14 at 23:44:12 UTC, the second started on 2022-02-25 at 04:38:09 UTC and ended on 2022-02-27 at 18:46:09 UTC. The average livetime for the three telescopes was 835719 seconds. Preliminary steps were to select the source against the background and to correct the arrival time of the events. The former task was achieved by selecting the source in the instrument field of view and identifying a region for background subtraction. This was done with the SAOIMAGEDS9 software (*45*). The source counts and the background were extracted from a circular region with radius $46''$ and a concentric annulus with inner and outer radius of $106''$ and $293''$, respectively (see Figure S1). We then converted photon arrival times to the Solar System Barycenter with the BARYCORR FTOOL included in HEASOFT 6.30.1, using the object coordinates in the FITS files, the DE421 JPL ephemeris and the ICRS reference frame.

Timing analysis

We searched around the known rotation frequency of the pulsar, $f = 0.115092$ Hz (*15*), and in a range of frequency derivatives with a $Z_n^2$ search (*46*). We used the quasi-fast folding algorithm included in the HENDRICS software v.7.0 (*47*, *48*), based on stingray 1.0 (*49*). We ran HENZSEARCH, initially using 16 bins for the pre-folding and $n = 1$ (sinusoidal pulsations). Once we determined a first solution around $0.115079$ Hz, we found that the pulse profile was better described by at least 5 harmonics using the H-test (*50*). Hence, we re-ran the $Z_n^2$ search, this time using $n = 5$ and 64 bins for pre-folding. Using the 90% confidence limits on the power (*51*, *52*) [as adapted by (*53*) and implemented in HENDRICS using the machinery in STINGRAY.STATS], we determined the 68.3% confidence limit on the frequency $f = 0.115079332(8)$ Hz and on the spin derivative of $\dot{f} = -2.1(7) \times 10^{-14}$ Hz s$^{-1}$. We verified the results independently in the following way: using HENPHASEOGRAM, we split the observation in 32 intervals and calculated the times of arrival (TOAs) of the pulsations in each interval with the FFTFIT algorithm (*54*); then, we used the PINT software (*55*) to fit these TOAs with a spin-down law, obtaining compatible values for both the frequency and the upper limit on the frequency derivative.

Spectral analysis

Spectral analysis was carried out with the python interface to the x-ray spectral-fitting program XSPEC, version 12.12.1 (*17*). Interstellar absorption was taken into account by using the XSPEC model TBABS with abundances from (*56*). Due to the limited energy resolution and range covered by *IXPE*, we tested only three simple models:

1. TBABS*(BBODY+POWERLAW), which is a standard spectral decomposition for 4U 0142+61 [see e.g. (*18*)],
2. TBABS*(BBODY+BBODY), which is another model often applied to magnetar sources, and
3. TBABS*(BBODY+TRCPOW), where TRCPOW is a truncated power-law, that is, a power-law which drops to zero below an energy threshold, $E_{\mathrm{trc}}$. This model is introduced to provide a better phenomenological representation of Resonant Compton Scattering [RCS, (*25, 26, 57*)] spectra by avoiding the unwarranted contribution of the power law at low energies. According to the RCS scenario, in fact, the power-law is produced by the up-scattering of thermal photons and is present only above an energy of typically a few keVs.

We found that the absorption is essentially unconstrained if we fit the *IXPE* spectrum alone. Therefore, we fixed the column density to $n_H = 0.57 \times 10^{22}$ cm$^{-2}$, taken from (*20*).

The spectra from the three *IXPE* telescopes were fitted simultaneously allowing for an energy-independent cross-normalization factor for the second and third detectors to account for the uncertainties in their absolute effective area calibration. The results of the background-subtracted spectral fitting in the energy interval 2–8 keV, which is the nominal *IXPE* range, are reported in Table S1 and shown in Figure S2. The fit with both the “standard” BBODY+POWERLAW and the BBODY+BBODY models are both acceptable, with a $\chi^2$ of 511.5 and 496.0, respectively, for 441 degrees of freedom. The fit parameters obtained for the BBODY+POWERLAW model (i.e. the blackbody temperature and the power-law photon index) are in broad agreement with those derived in previous observations (*18*), as well as the resulting unabsorbed flux, which for detector 1 (taken as a reference) turns out to be $\sim 6.5 \times 10^{-11}$ erg s$^{-1}$ cm$^{-2}$ in the 2–10 keV energy range. Cross-normalization factors are within 15%, which is usual for x-ray missions, especially in their early phases like *IXPE*.

The fit with the BBODY+TRCPOW suffers from residuals at low energy, that can be in large part removed by letting the column density free to vary. However, since the main goal of the present work is that of discussing the results of the *IXPE* polarization measurement of 4U 0142+61, we leave a further, more detailed analysis to future investigations, and use this spectral model with the purpose of performing a joint spectro-polarimetric analysis of the observed data.

We performed a preliminary phase-resolved spectral analysis, grouping the data in six phase bins; because of the too few counts at high energies we considered only the 2–7 keV range. The obtained spectra are shown in Figure S3, where the selected phases are also reported, together with the best fitting parameters for the $f$*TBABS*(BBODY+POWERLAW). No statistically significant spectral changes with phase were detected at this stage and we defer a more detailed analysis to a future work.

Polarization analysis

Polarization analysis was carried out with two independent approaches. The first uses the tools available in the IXPEOBSSIM simulation and analysis package, developed by the *IXPE* collaboration to generate realistic simulated observations and to process them (*21*). This model-independent polarization analysis is based on the unbinned procedure described in (*42*). While a weighted analysis can provide an increase in sensitivity (*58, 59*), at the time of writing a proper tool to perform it in a way that can be reproduced also outside the collaboration is not yet available. To ensure replicability of our analysis, here we opted for a simpler, more consolidated unweighted procedure. Stokes parameters are calculated event-by-event from the photoelectron track emission angle and calibrated for the known spurious modulation of the instrument (*60*). Since the Stokes parameters are additive (*42*), those referring to a given energy band are obtained by simply summing the parameters of all the events in the energy range of interest. The background was removed by subtracting each of the Stokes parameters from those of the source, as it is common practice for spectra.

XSPEC analysis was performed following (*61*). We build binned spectra for the Stokes parameters $I$, $Q$ and $U$, which are then fitted with the usual XSPEC procedure of *forward folding*. The instrumental response to polarization is accounted for with the modulation response function, provided as a part of the *IXPE* calibration database. The background is removed by subtracting from the source Stokes spectra the corresponding background ones, after rescaling the latter to the area of the selection region. Polarization in a given energy interval was calculated by assuming the TBABS*(BBODY+POWERLAW) model presented above and convolving it with the constant polarization model POLCONST provided by XSPEC.

The results of the independent analysis of the three *IXPE* detectors and those obtained joining them using the model-independent and XSPEC approaches are reported in Table S2 for different energy bands. Statistical uncertainties are associated to the measured polarization degree (PD) and angle (PA) according to standard practices. In Table S2 we report the $1\sigma$ (68.3% confidence level) uncertainty on both PD or PA calculated following (*42*), in the assumption that Stokes parameters are normally distributed and correlated, and that PD and PA are independent. When the measured PD is lower than the Minimum Detectable Polarization (MDP) at the 99% confidence level (*62*), we list in the table both the best-estimate value and the MDP of the measurement; for PA, instead, we report the best-estimate value and extend the uncertainty interval over the entire range, $\mathrm{PA}-90°$–$\mathrm{PA}+90°$. Uncertainties for the XSPEC analysis are derived with the ERROR command of XSPEC and one parameter of interest.

Actually, it is well known that the polarization degree and angle are not independent (especially for low-significance measurements). For this reason, we show also the regions representing the 68.3% confidence level for the joint measurement of the polarization degree and angle (see Figure 2). In the IXPEOBSSIM analysis, this is derived with standard functions (*62*–*64*) from the measured quantities. This approach assumes that the Stokes parameters are normally distributed and uncorrelated, which is a good approximation for the polarization degree exhibited by 4U 0142+61 and for the *IXPE* modulation factor in the 2–8 keV energy range. For the XSPEC analysis, we used the STEPPAR command to draw the contour region, assuming two parameters of interest.

We also performed a joint spectro-polarimetric analysis with XSPEC, fitting simultaneously all the Stokes spectra $I$, $Q$ and $U$ with models that include both spectral and polarization information as a function of the energy. In particular, we started from the three models discussed above for the spectral analysis and associated to each additive component a different polarization, assumed to be constant with energy. The column density and cross-calibration of the three detectors were frozen at the same values of spectral analysis (see Table S1), whereas all the other parameters are left free to vary.

Fit results are shown in Figures S4–S6 and parameters are reported in Table S3. They show some remarkable features. First of all, the spectral parameters are in agreement, within statistical uncertainties, with the corresponding ones obtained in spectral analysis alone: this suggests that the adopted spectral decomposition is consistent with polarization data and points towards a likely different physical origin for the two components, each characterized by its own spectral and polarization properties. The polarization degree of the two components is strongly model dependent; for example, the low-energy blackbody has a polarization of $\sim 60\%$ when the model comprises also a power-law, and of $\sim 17\%$ when the second component is a truncated power law. Such a large variation is easy to explain as a consequence of the fact that the two models imply a different contribution of the high-energy component at low energies and therefore their polarization must adjust so that their sum is the observed polarization. The results obtained with the TBABS*(BBODY+TRCPOW) model are free of this issue and so are more trustworthy. Nonetheless, essentially in all cases we do observe a swing of the polarization angle of $\sim 90°$ between the two components, which is therefore a robust feature of spectro-polarimetric fitting.

<u>Rotating vector model</u>

The rotating vector model [RVM; (*23*), see also (*65*)] provides a simple way for computing the polarization angle of radiation coming from a small (point-like) region located at (or close to) the magnetic pole of a neutron star. In fact, if the star magnetic field is a dipole so that $\boldsymbol{B}$ at the emitting point is along the magnetic axis, the angle between the projection of the field in the plane of the sky and a reference direction, taken as the projection in the same plane of the star spin axis, is given by (*65*)

$$\tan\alpha = \frac{\sin\xi\sin\gamma}{\cos\chi\sin\xi\cos\gamma - \sin\chi\cos\xi} \quad (1)$$

where $\chi$ and $\xi$ are the inclinations of the observer's line-of-sight and the star dipole axis with respect to the spin axis, and $\gamma$ is the rotational phase. The angle $\alpha$ coincides with the polarization angle PA.

In case radiation comes from an extended region, the situation is more complicated because $\alpha$ is different at the different emission points, since the direction of $\boldsymbol{B}$ changes substantially on the surface. It can be shown that now PA is given by (*10*)

$$\tan(2\mathrm{PA}) = \frac{\Sigma_i^{N_X}\sin(2\alpha_i) - \Sigma_i^{N_O}\sin(2\alpha_i)}{\Sigma_i^{N_O}\cos(2\alpha_i) - \Sigma_i^{N_X}\cos(2\alpha_i)}, \quad (2)$$

where the two summations are taken over the total number of X (O) mode photons and

$$\tan\alpha_i = -\frac{B_{y,i}}{B_{x,i}}; \quad (3)$$

here $B_{y,i}$, $B_{x,i}$ are the (cartesian) components of the (local) magnetic field in the plane of the sky.

It is easy to check that if all the $\alpha_i$ are equal to the same value, $\alpha$, we still get $\mathrm{PA} = \alpha$. While this does not hold in general for emission coming from the surface, it becomes true if the polarization direction is determined by the properties on a sphere of radius much larger than the star radius. In fact, here the dipolar field direction changes little from point to point, so that $\tan\alpha_i$ is about the same for all photons. Moreover, the common value $\alpha$ turns out to be given again by equation (1). In fact, the cartesian components of the field perpendicular to the line-of-sight are related to the those referred to the dipole axis by

$$B_x = B_p p_x + B_q q_x + B_t t_x$$
$$B_y = B_p p_y + B_q q_y + B_t t_y \quad (4)$$

where $\boldsymbol{p}$, $\boldsymbol{q}$ and $\boldsymbol{t}$ are the axes unit vectors, with $\boldsymbol{t}$ along the dipole axis. By relating $B_p$, $B_q$ and $B_t$ to the polar components of the dipole field and summing over the entire surface, it can be seen that only the $t$ contribution survives in equations (4), with $t_y$ and $t_x$ given exactly by the numerator and denominator of equation (1) [see (*10*)].

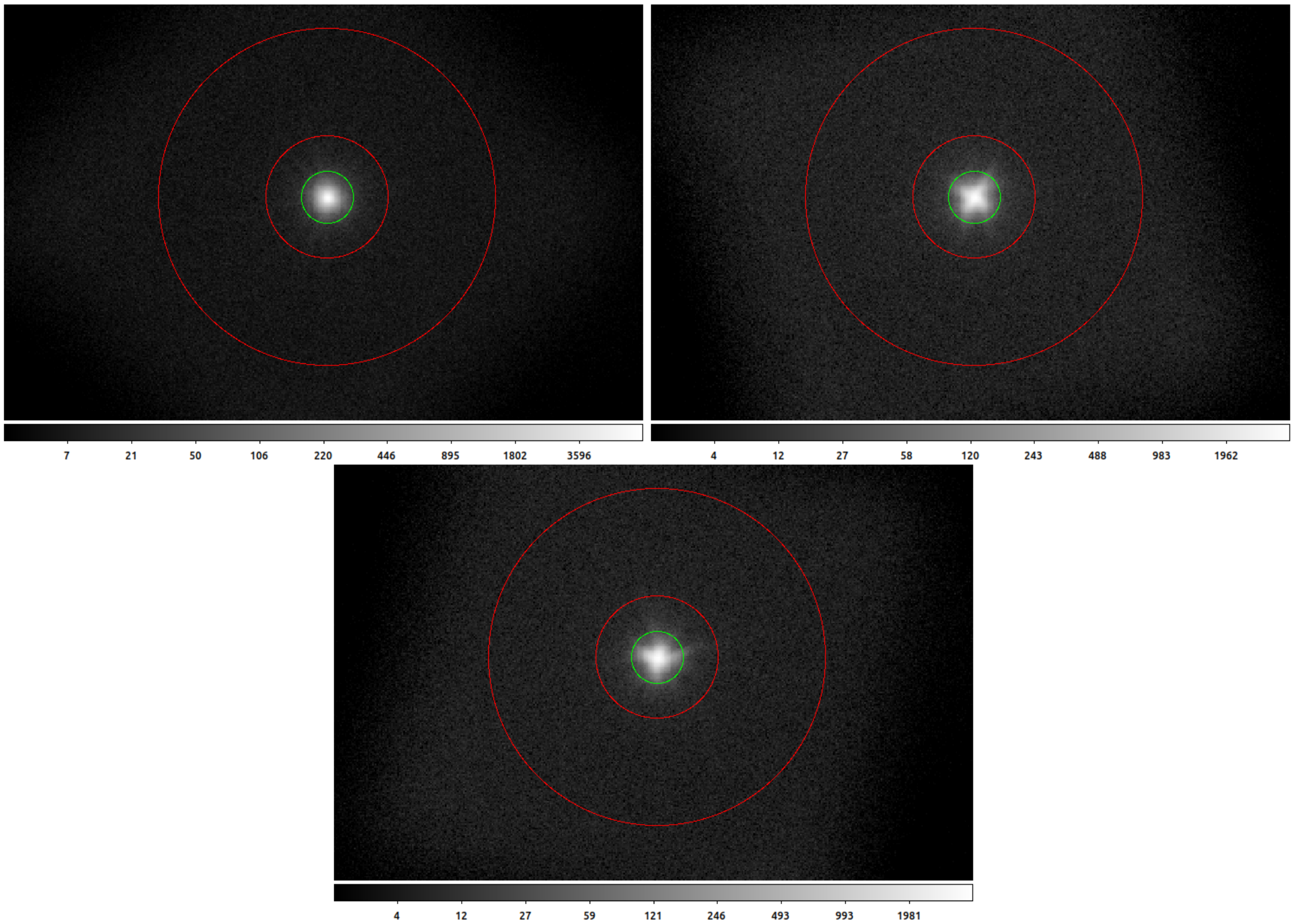

**Fig. S1.**
Regions selected with SAOImageDS9. The green circle and the red annulus mark the regions selected for extracting the source and background counts, respectively. The gray scale is logarithmic to highlight the background.

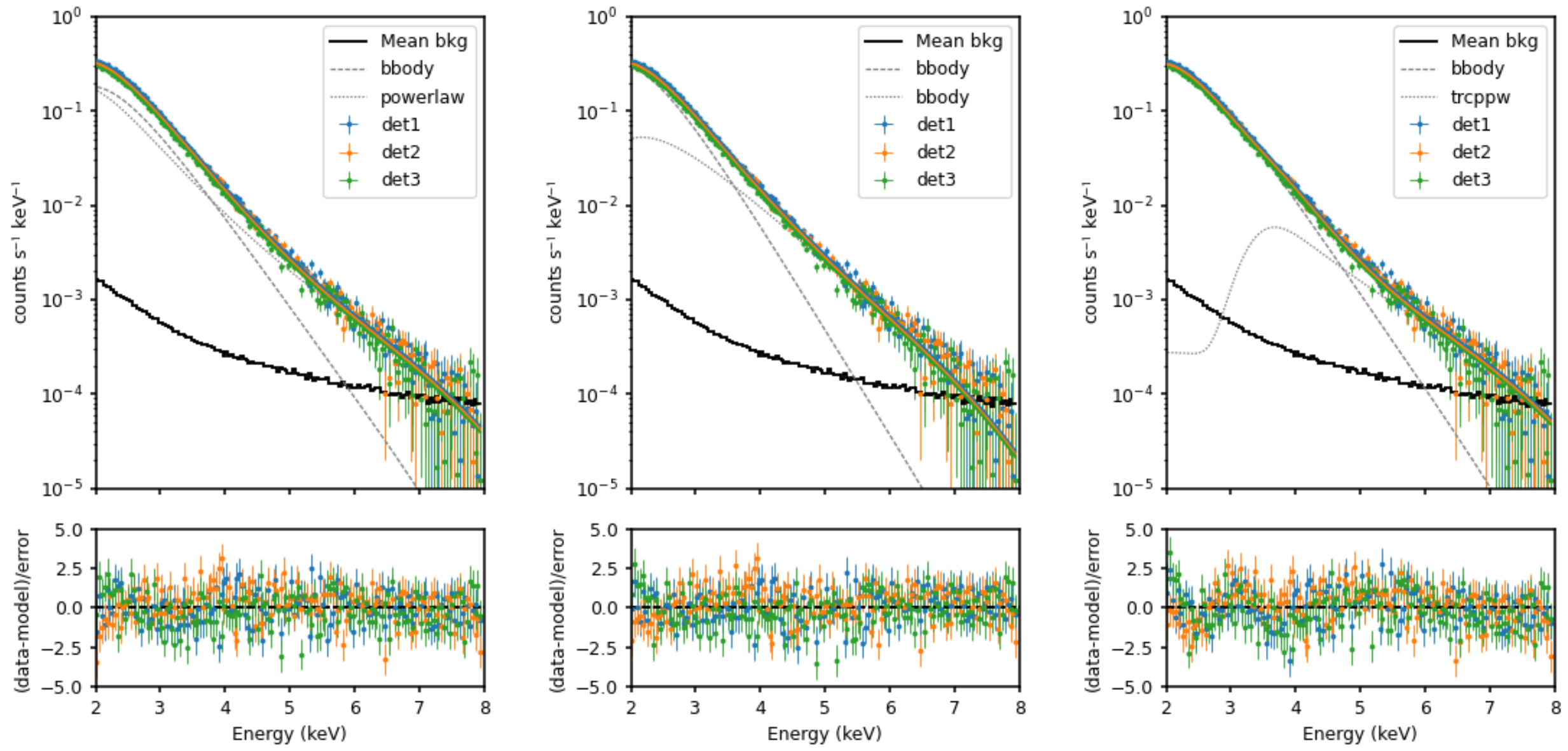

**Fig. S2.**

Spectral fitting for the three IXPE detectors. The TBABS*(BBODY+POWERLAW) is on the left, TBABS*(BBODY+BBODY) at the center and TBABS*(BBODY+TRCPOW) on the right. The additive components of each model and the average background counts are also shown.

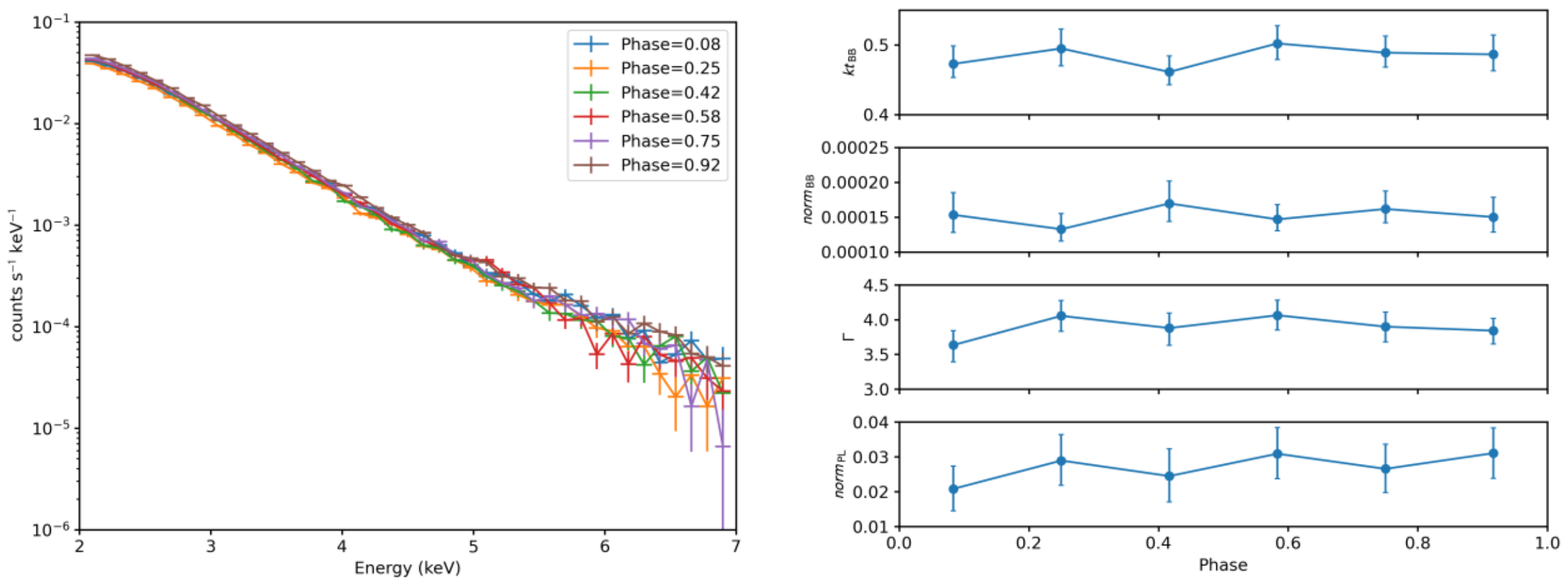

**Fig. S3.**

Phase-resolved spectral analysis of 4U 0142+61. Left panel: spectra obtained by folding the data at the measured spin period and grouping them in six, equally-spaced phase intervals. Right panel: phase-dependent behavior of the spectral parameters obtained fitting the spectrum in each phase bin with the *f* *TBABS*(BBODY+POWERLAW) model (see Figure S2, left panel). Phases refer to the center of each bin.

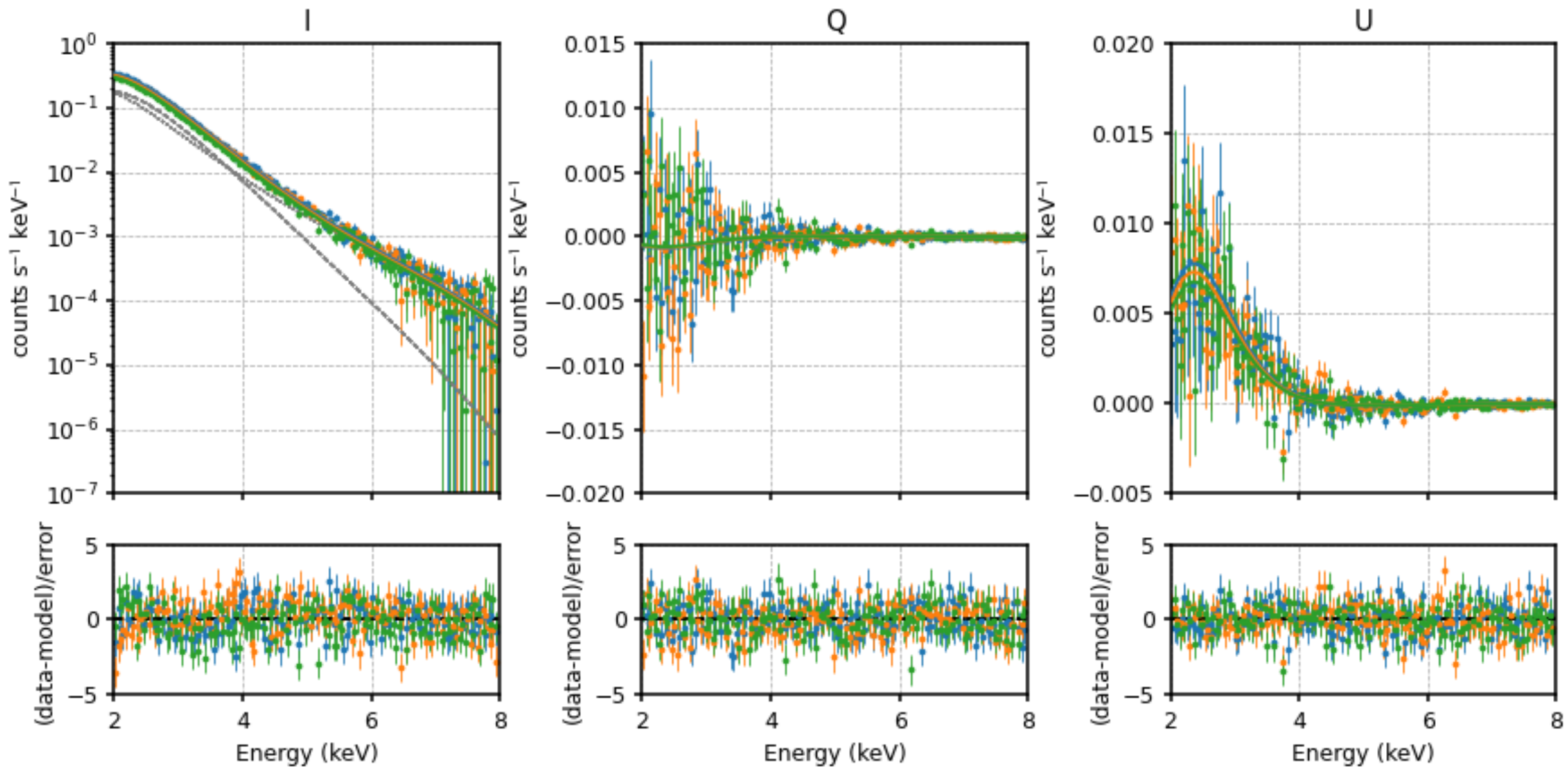

**Fig. S4.**

Fit of $I$, $Q$ and $U$ with the TBABS*(BBODY*POLCONST+POWERLAW*POLCONST) model (see Table S3 for fit parameters).

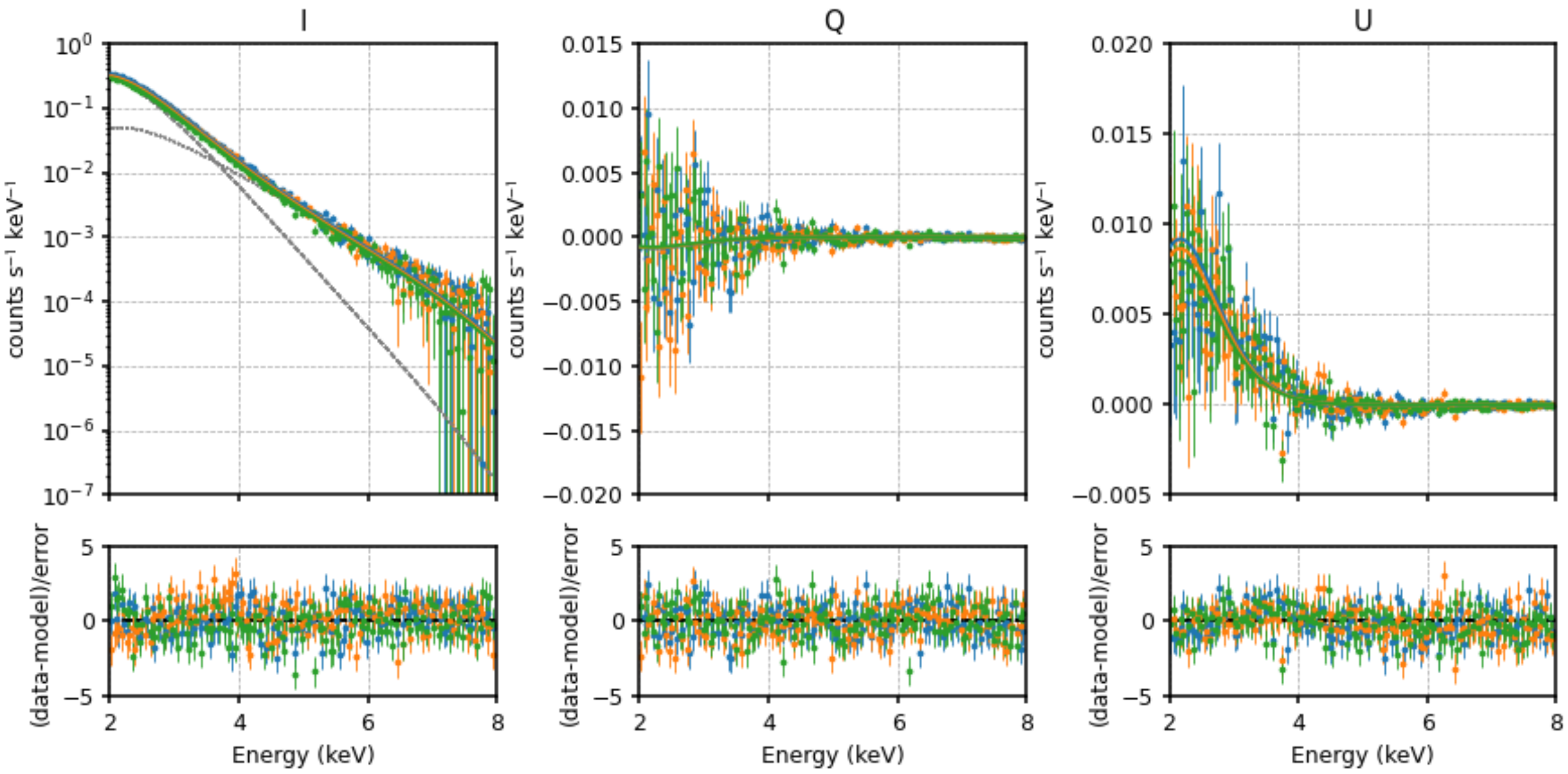

**Fig. S5.**

Same as Figure S4, but for the TBABS*(BBODY*POLCONST+BBODY*POLCONST) model.

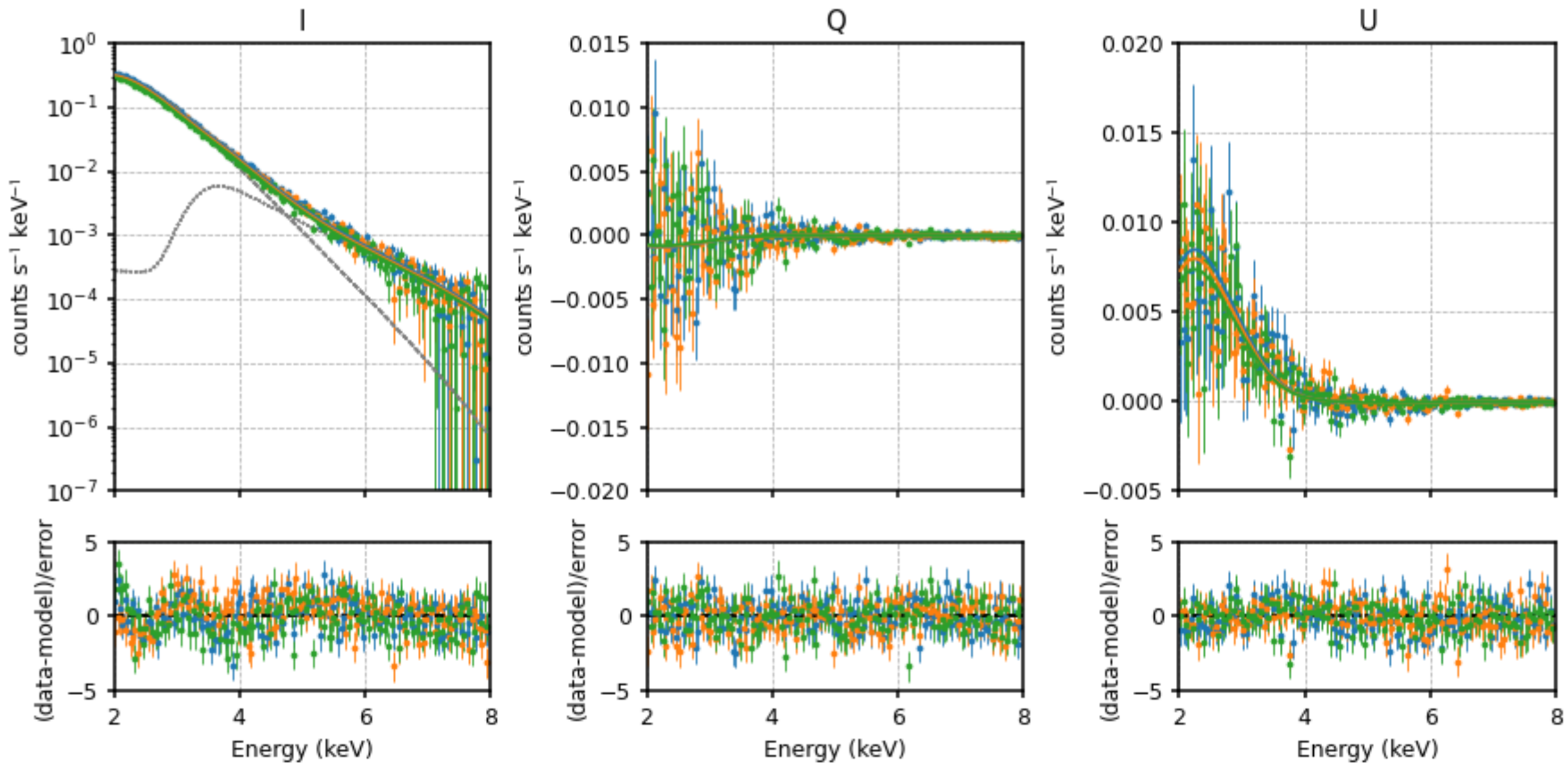

**Fig. S6.**
Same as Figure S4, but for the TBABS*(BBODY*POLCONST+TRCPOW*POLCONST) model.

| $f$*TBABS*(BB+PL) | | $f$*TBABS*(BB$_1$+BB$_2$) | | $f$*TBABS*(BB+TPL) | |
|---|---|---|---|---|---|
| $n_H$[a] | 0.57 | $n_H$ | 0.57 | $n_H$ | 0.57 |
| $kT_{\rm BB}$[b] | $0.471^{+0.004}_{-0.004}$ | $kT_{\rm BB1}$ | $0.399^{+0.004}_{-0.004}$ | $kT_{\rm BB}$ | $0.4470^{+0.0006}_{-0.0009}$ |
| $\rm norm_{BB}$[c] | $0.00108^{+0.00004}_{-0.00003}$ | $\rm norm_{BB1}$ | $0.00196^{+0.00001}_{-0.00001}$ | $\rm norm_{BB}$ | $0.002073^{+0.000007}_{-0.000007}$ |
| $\Gamma$[d] | $3.69^{+0.05}_{-0.05}$ | $kT_{\rm BB2}$ | $0.81^{+0.02}_{-0.02}$ | $\Gamma$ | $2.69^{+0.06}_{-0.04}$ |
| $\rm norm_{PL}$[e] | $0.119^{+0.008}_{-0.008}$ | $\rm norm_{BB2}$ | $0.00040^{+0.00002}_{-0.00002}$ | $\rm norm_{PL}$ | $0.028^{+0.001}_{-0.002}$ |
| | | | | $E_{\rm trc}$[f] | $3.340^{+0.058}_{-0.002}$ |
| $f_{\rm det1}$[g] | 1.0 | $f_{\rm det1}$ | 1.0 | $f_{\rm det1}$ | 1.0 |
| $f_{\rm det2}$[h] | $0.963^{+0.003}_{-0.003}$ | $f_{\rm det2}$ | $0.963^{+0.003}_{-0.003}$ | $f_{\rm det2}$ | $0.963^{+0.003}_{-0.003}$ |
| $f_{\rm det3}$[i] | $0.855^{+0.003}_{-0.003}$ | $f_{\rm det3}$ | $0.855^{+0.003}_{-0.003}$ | $f_{\rm det3}$ | $0.855^{+0.002}_{-0.002}$ |
| $\chi^2$ | 511.5 with 441 dof | $\chi^2$ | 496.0 with 441 dof | $\chi^2$ | 586.2 with 440 dof |

**Table S1.**

Results of the spectral fitting of the *IXPE* data. A constant factor $f$ is assumed for accounting of *IXPE* detectors mutual cross-calibration. Uncertainties are calculated at the 68.3% confidence level.

[a] Column density in units of $10^{22}$ cm$^{-2}$ (frozen parameter).
[b] Blackbody temperature in keV.
[c] Blackbody normalization, $L_{39}/D_{10}^2$, with $L_{39}$ the luminosity in units of $10^{39}$ erg s$^{-1}$ and $D_{10}$ the distance in units of 10 kpc.
[d] Power-law photon index.
[e] Power-law normalization in photons keV$^{-1}$ cm$^{-2}$ s$^{-1}$ at 1 keV.
[f] Truncated power-law energy threshold in keV.
[g] Constant cross-calibration factor for detector 1 (frozen parameter).
[h] Constant cross-calibration factor for detector 2.
[i] Constant cross-calibration factor for detector 3.

| | 2-3 keV | 3-4 keV | 4-4.75 keV | 4.75-5.5 keV | 5.5-8 keV | 2-8 keV |
|---|---|---|---|---|---|---|
| PD - det1 [%] | $13^{+2}_{-2}$ | $15^{+2}_{-2}$ | 4 (15) | 16 (25) | 25 (34) | $11^{+1}_{-1}$ |
| PD - det2 [%] | $15^{+2}_{-2}$ | $13^{+2}_{-2}$ | 13 (15) | 7 (26) | $46^{+11}_{-11}$ | $13^{+1}_{-1}$ |
| PD - det3 [%] | $16^{+2}_{-2}$ | $13^{+2}_{-2}$ | 3 (16) | 17 (28) | $53^{+13}_{-13}$ | $12^{+1}_{-1}$ |
| PD - sum [%] | $14^{+1}_{-1}$ | $13^{+1}_{-1}$ | 5 (9) | 11 (15) | $41^{+7}_{-7}$ | $12^{+1}_{-1}$ |
| PD – sum (XSPEC) [%] | $16^{+1}_{-1}$ | $14^{+1}_{-1}$ | 6 (9) | 10 (15) | $35^{+7}_{-7}$ | $13^{+1}_{-1}$ |
| PA - det1 [deg] | $46^{+3}_{-3}$ | $48^{+4}_{-4}$ | $-15^{+105}_{-75}$ | $-33^{+123}_{-57}$ | $-47^{+137}_{-43}$ | $46^{+3}_{-3}$ |
| PA - det2 [deg] | $52^{+3}_{-3}$ | $54^{+5}_{-5}$ | $47^{+43}_{-137}$ | $-71^{+161}_{-19}$ | $-44^{+7}_{-7}$ | $53^{+3}_{-3}$ |
| PA - det3 [deg] | $42^{+3}_{-3}$ | $57^{+5}_{-5}$ | $34^{+56}_{-124}$ | $-70^{+160}_{-20}$ | $-42^{+7}_{-7}$ | $45^{+3}_{-3}$ |
| PA - sum [deg] | $47^{+2}_{-2}$ | $52^{+3}_{-3}$ | $37^{+53}_{-127}$ | $-54^{+144}_{-36}$ | $-44^{+5}_{-5}$ | $48^{+2}_{-2}$ |
| PA – sum (XSPEC) [deg] | $47^{+2}_{-2}$ | $53^{+3}_{-3}$ | $41^{+49}_{-131}$ | $-64^{+154}_{-26}$ | $-39^{+6}_{-6}$ | $48^{+2}_{-2}$ |

**Table S2.**

Table summarizing: (i) the polarization degree and angle measured independently with the three *IXPE* detectors and model-independent analysis (*42*), as implemented in IXPEOBSSIM; (ii) the value obtained summing the data from the three *IXPE* telescopes with the same approach and (iii) the value obtained with the summed data but with the XSPEC procedure. All the data sets provide statistically compatible values. Uncertainties are calculated for a 68.3% confidence level, assuming that the polarization degree and angle are independent. When the measured value of the polarization degree is lower than the Minimum Detectable Polarization (MDP) at the 99% confidence level, we show the latter in parenthesis and assume that the polarization angle can vary in its maximum interval.

| $f$*TBABS*(BB*POLC$_1$+PL*POLC$_2$) | | $f$*TBABS*(BB$_1$*POLC$_1$+BB$_2$*POLC$_2$) | | $f$*TBABS*(BB*POLC$_1$+TPL*POLC$_2$) | |
|---|---|---|---|---|---|
| $n_{\mathrm{H}}$[a] | 0.57 | $n_{\mathrm{H}}$ | 0.57 | $n_{\mathrm{H}}$ | 0.57 |
| $kT_{\mathrm{BB}}$[b] | $0.473^{+0.004}_{-0.004}$ | $kT_{\mathrm{BB1}}$ | $0.401^{+0.003}_{-0.004}$ | $kT_{\mathrm{BB}}$ | $0.448^{+0.001}_{-0.002}$ |
| $\mathrm{norm}_{\mathrm{BB}}$[c] | $0.00106^{+0.00003}_{-0.00003}$ | $\mathrm{norm}_{\mathrm{BB1}}$ | $0.00196^{+0.00001}_{-0.00001}$ | $\mathrm{norm}_{\mathrm{BB}}$ | $0.002073^{+0.000005}_{-0.000013}$ |
| $\mathrm{PD}_1$[d] | $0.59^{+0.06}_{-0.06}$ | $\mathrm{PD}_1$ | $0.23^{+0.02}_{-0.02}$ | $\mathrm{PD}_1$ | $0.168^{+0.008}_{-0.008}$ |
| $\mathrm{PA}_1$[e] | $48^{+2}_{-2}$ | $\mathrm{PA}_1$ | $47^{+2}_{-2}$ | $\mathrm{PA}_1$ | $48^{+1}_{-1}$ |
| $\Gamma$[f] | $3.71^{+0.04}_{-0.05}$ | $kT_{\mathrm{BB2}}$ | $0.82^{+0.02}_{-0.02}$ | $\Gamma$ | $2.67^{+0.04}_{-0.07}$ |
| $\mathrm{norm}_{\mathrm{PL}}$[g] | $0.123^{+0.008}_{-0.008}$ | $\mathrm{norm}_{\mathrm{BB2}}$ | $0.00038^{+0.00002}_{-0.00002}$ | $\mathrm{norm}_{\mathrm{PL}}$ | $0.027^{+0.002}_{-0.002}$ |
| | | | | $E_{\mathrm{trc}}$[h] | $3.38^{+0.04}_{-0.02}$ |
| $\mathrm{PD}_2$[i] | $0.39^{+0.06}_{-0.06}$ | $\mathrm{PD}_2$ | $0.06^{+0.03}_{-0.03}$ | $\mathrm{PD}_2$ | $0.23^{+0.04}_{-0.04}$ |
| $\mathrm{PA}_2$[j] | $-42^{+4}_{-4}$ | $\mathrm{PA}_2$ | $-51^{+16}_{-16}$ | $\mathrm{PA}_2$ | $-44^{+5}_{-5}$ |
| $f_{\mathrm{det1}}$[k] | 1.0 | $f_{\mathrm{det1}}$ | 1.0 | $f_{\mathrm{det1}}$ | 1.0 |
| $f_{\mathrm{det2}}$[l] | 0.963 | $f_{\mathrm{det2}}$ | 0.963 | $f_{\mathrm{det2}}$ | 0.963 |
| $f_{\mathrm{det3}}$[m] | 0.855 | $f_{\mathrm{det3}}$ | 0.855 | $f_{\mathrm{det3}}$ | 0.855 |
| $\chi^2$ | 1337.8 with 1333 dof | $\chi^2$ | 1360.4 with 1333 dof | $\chi^2$ | 1425.6 with 1332 dof |

**Table S3.**

Results of the spectro-polarimetric fitting with the same models used for spectral analysis in Table S1. The detector cross-calibration factors and the column density are frozen to the values found in the spectral analysis. All the other parameters are compatible with the values found from the spectral fitting only. Uncertainties are calculated as the extrema of the 68.3% confidence contours.

[a] Column density in units of $10^{22}$ cm$^{-2}$ (frozen parameter).

[b] Blackbody temperature in keV.

[c] Blackbody normalization, $L_{39}/D_{10}^2$, with $L_{39}$ the luminosity in units of $10^{39}$ erg s$^{-1}$ and $D_{10}$ the distance in units of 10 kpc.

[d] Polarization degree of the first spectral component.

[e] Polarization angle of the first spectral component in deg.

[f] Power-law photon index.

[g] Power-law normalization in photons keV$^{-1}$ cm$^{-2}$ s$^{-1}$ at 1 keV.

[h] Truncated power-law energy threshold in keV.

[i] Polarization degree of the second spectral component.

[j] Polarization angle of the second spectral component in deg.

[k] Constant cross-calibration factor for detector 1 (frozen parameter).

[l] Constant cross-calibration factor for detector 2 (frozen parameter).
[m] Constant cross-calibration factor for detector 3 (frozen parameter).